\documentclass[12pt]{article}
\usepackage{epsfig}
\usepackage{amsmath}
\usepackage{amssymb}
\usepackage{subfigure}
\usepackage{graphicx}
\usepackage{verbatim}

\begin{document}

\begin{titlepage}
\begin{center}

{\Large Discrete Symmetries of Chaotic Strings}

\vspace{2.cm} {\bf Mirko Sch\"afer}

\vspace{0.5cm}
Frankfurt Institute of Advanced Studies,
Ruth-Moufang-Str.1,

60438 Frankfurt, Germany

\vspace{2.cm} {\bf Christian Beck}

\vspace{0.5cm}

School of Mathematical Sciences, Queen Mary, University
of London,

Mile End Road, London E1 4NS, UK

\vspace{2cm}

\end{center}

\abstract{Chaotic strings are particular classes of coupled map
lattices that can serve as models for vacuum fluctuations in
stochastically quantized field theories.
 They have been
previously shown to distinguish standard model coupling
parameters as corresponding to states of strongest possible
chaotic behaviour and vanishing nearest-neighbour correlation. In
this paper we look at discrete symmetry transformations for
chaotic strings. We show that several of the chaotic string
dynamics can be transformed into each other by simple discrete
coordinate transformations.
We investigate how
expectation values converge in the various coupling parameter
regions and single out those stable zeros of the correlation
function that correspond to ergodic states with well-defined
convergence properties.}


\vspace{1.3cm}

\end{titlepage}

\newpage

\section{Introduction}
Coupled map lattices (CMLs) as introduced by Kaneko and Kapral
\cite{kaneko,kapral} some 24 years ago are a paradigm of
spatially extended dynamical systems with discrete evolution in
space and time. They exhibit a rich structure of complex
dynamical phenomena
\cite{kanekobook,kanekobook2,Beck:2002,Bunimovich:1988,amritkar,amritkar2,
carretero} .
Of particular interest are CMLs that exhibit spatio-temporal
chaotic behaviour. In certain cases it can be rigorously proved
that the system possesses an ergodic invariant measure
\cite{bunimovich,baladi,Bricmont:1996,Jarvenpaa:2001, Baladi:2000} provided the
coupling is small enough. Various interesting aspects of chaotic
CMLs have been investigated in
\cite{daido,chate,ding,mackey,Keller:1992,Lemaitre:1997,Chate:1997,groote,groote2}.

CMLs are not only of theoretical interest in the theory of
dynamical systems but they also have applications for a variety of
physical problems, e.g.\ as models for hydrodynamical flows,
turbulence, chemical reactions, or synchronization. A very
interesting new application in quantum field theories has been
pointed out in \cite{Beck:2002, physica-d}: Certain types
of coupled map lattices can serve as models of vacuum fluctuations.
This may at first
sight look as a rather exotic type of theory. The remarkable
observation, however, is that these types of coupled map lattices
seem to distinguish certain numerical values of coupling
constants that coincide with those observed in the standard model
of electroweak and strong interaction. For more details on this,
see \cite{Beck:2002, physica-d, Beck:1995, pla, dark, beck-recent}. The main idea underlying
this approach is that indeed the fundamental constants of nature
may be fixed by dynamically evolving chaotic states of vacuum
fluctuations that possess the strongest random
properties that are possible for a smooth coupled deterministic
chaotic system. These types of vacuum fluctuations may underly
the currently observed dark energy in the universe \cite{dark,
phantom}.

The relevant classes of coupled map lattices for this approach
have been termed `chaotic strings' \cite{Beck:2002, physica-d}.
This is a name only, it should be clear that chaotic strings are
very different from `ordinary' strings studied in string theory.
Chaotic strings are closely related to an alternative way of
quantizing classical systems which is based on the so-called
`chaotic quantization' method \cite{Beck:1995, biro, maher}.
Chaotic quantization yields additional sectors of
highly nonlinear field theories, which
can be used to stabilize the fundamental constants of nature. In
theories of quantum gravity standard model parameters are often
thought of as being related to certain types of scalar fields,
so-called moduli fields \cite{beck-recent, gardner}. A varying standard model parameter (e.g.
the fine structure constant) can be essentially regarded as such
a moduli field. These moduli fields evolve to minima of their
potentials. In this way for example the fine structure constant
stabilizes at its currently observed low-energy value of
$1/137.036$. The main idea underlying chaotic quantization methods
is that the chaotic string dynamics produces suitable effective
potentials for moduli fields which fix and stabilize fundamental
constants, avoiding anthropic considerations.

Chaotic strings can occur in different versions, depending on the
way the coupling is done. In this paper, we study discrete
symmetry transformations for the various possible chaotic string
dynamics. From a nonlinear dynamics point of view, chaotic
strings are just diffusively or anti-diffusively coupled map
lattices consisting locally of positive or negative $N$-th order
Tchebyscheff maps $(N\geq 2)$
\cite{Beck:2002,groote,groote2,Beck:1993,higher1,higher2,dettmann1,dettmann2}
with either forward or backward coupling. These live on
1-dimensional lattices with periodic boundary conditions. For each
$N$ there are {\em a priori} 8 different chaotic strings. But we
will show that this actually reduces to four different relevant
strings if $N$ is even and two different relevant strings if $N$
is odd, due to the fact that there is a number of discrete
coordinate transformations which one can use to map one string
dynamics into another one. Our paper for the first time
investigates these discrete symmetries in a systematic and
complete way. In ordinary quantum field theories, it is well
known that discrete symmetries such as CPT (charge, parity, time
reversal) play an important role. Our paper here investigates the
analogue of this for chaotic strings.

Our main examples are chaotic strings build up from Tchebyscheff
maps of order $N=2$ and $N=3$, which can be regarded as kind of
`ground states' (with even or odd parity) in quantum gravity
embeddings of chaotic strings \cite{Beck:2002}. For the
calculation of expectations of observables associated with the
chaotic \string dynamics it is important to know whether the
dynamics is ergodic. In fact, non-ergodicity can be related to
spontaneous symmetry breaking in this context (a particular
attractor out of several is spontaneously chosen, depending on
the inital condition). For more details on this symmetry breaking
aspect, see \cite{pla}. To completely understand the discrete
symmetry transformation behaviour of chaotic strings it is thus
important to investigate the ergodic properties and the
convergence behaviour of observables in the various parameter
regions. We will investigate this question numerically and define
some suitable measures of homogenity, which test the dependence
of expectations of local observables under translations on the
lattice and under changes of the initial conditions. An
observable of particular interest is the nearest neighbour
correlation function, whose zeros were used in \cite{Beck:2002,
physica-d} to  single out distinguished standard model
parameters. Our investigation here analyses which of these zeros
are in parameter regions with ergodic behaviour and well-defined
convergence properties of expectations.
Interesting enough, our analysis
shows that precisely those zeros that correspond to ergodic
behaviour are those for which a straight-forward standard model
parameter interpretation has been previously found in \cite{Beck:2002,
physica-d}. All this emphasizes the importance of fully
understanding the transformation behaviour of chaotic strings
under discrete symmetry operations.


This paper is organized as follows. In section 2 we define the
chaotic string dynamics. The invariance under discrete
symmetry transformations is systematically investigated in section
3 (detailed calculations in the appendix).
The transformation behaviour of certain relevant observables of
the chaotic strings (such as the self energy and the
interaction energy) is discussed in section 4.
Finally, in section 5 we present our numerical results on the
ergodic behaviour.

\section{Definition}

A chaotic string is a dynamical system defined on a chain of $J$
lattice points with periodic boundary conditions. The dynamical
variable (or phase)
 at lattice site $i$ at time step $t$ is denoted by $\Phi_{t}^{i}$.
 The right-hand neighbour of the $J$th site is the first lattice point, i.e. $\Phi^{J+1}=\Phi^{1}$
 and the left-hand neighbour of the first site is the $J$th lattice point,
 i.e. $\Phi^{0}=\Phi^{J}$.

The $J$ dynamical variables $\Phi_t^i$ form a $J$-dimensional
vector $\mathbf{\Phi}_t\in S$. The phase space $S\subset
\mathbb{R}^{J}$ is given by
\begin{equation}
 S=\left\{\mathbf{r}=\sum_{i=1}^{J}x^{i}\mathbf{e}_{i}\ :\ \ |x^{i}|\leq 1 \forall\ \  i=1,2,...,J\right\},
\end{equation}
where $\{\mathbf{e}_1,...,\mathbf{e}_{J}\}$ can be chosen as the
standard Cartesian basis of $\mathbb{R}^{J}$.

The time evolution of the lattice dynamical system corresponds to
iterating a vector-valued function $\mathbf{f}: S \to S$. We
denote
\begin{equation}
\mathbf{f}^{n}=\underbrace{\mathbf{f}\circ\mathbf{f}\circ\ldots\circ\mathbf{f}}_{n\mbox{ times}}
\end{equation}
unless otherwise stated.

The chaotic string dynamics studied in \cite{Beck:2002} evolve
according to

\begin{equation}
 \Phi_{t+1}^{i} =
(1-a)T_{N}(\Phi_{t}^{i})+s\frac{a}{2}\left(T^{b}_{N}(\Phi_{t}^{i-1})+T^{b}_{N}
(\Phi_{t}^{i+1})\right),\\
\label{eq:beck_dyn}
\end{equation}

where $T_{N}(\Phi)$ is a Tchebyscheff polynomial of order $N$.
Tchebyscheff maps are conjugated to a Bernoulli shift (see, e.g., \cite{Beck:1993})
and are distinguished by minimum higher-order corelations \cite{higher1, higher2}.
One
has
\begin{eqnarray}
 T_{1} (\Phi)&=&\Phi,\\
 T_{2}(\Phi) &=& 2\Phi^{2}-1,\\
 T_{3}(\Phi) &=& 4\Phi^{3}-3\Phi.
 \end{eqnarray}
Tchebyscheff polynomials of higher order $N>3$ are recursively
defined by
\begin{equation}
T_{N+1}(\Phi) = 2\Phi\cdot T_{N}(\Phi)-T_{N-1}(\Phi).
\end{equation}

In the language of nonlinear dynamics a chaotic string is just a 1-dimensional coupled map
lattice (CML) of Tchebyscheff maps with nearest-neighbour
coupling \cite{physica-d, dettmann1, dettmann2}.
We are also considering the negative polynomials $T_{-N}(\Phi):=-T_{N}(\Phi)$. The coupling constant $a$ is taken in the interval~$[0,1]$, the value of $s$ is chosen as~``$+$'' (diffusive coupling) or ``$-$'' (anti-diffusive coupling). Finally the integer~$b=1$ accounts for forward coupling $T^{1}_{N}(\Phi)=T_{N}(\Phi)$, whereas $b=0$ stands for backward coupling with $T^{0}_{N}(\Phi):=\Phi$.\\
For each $N$ the different combinations $(\pm N,s,b)$ lead to $8$ different evolution equations, which can be written in the form
\begin{equation}
 \mathbf{\Phi}_{t+1}=\mathbf{f}_{[N,s,b]}(\mathbf{\Phi}_t;a)
\end{equation}
with
\begin{equation}
 f_{[N,s,b]}^{i}(\mathbf{\Phi};a)=(1-a)T_{N}(\Phi^{i})+s\frac{a}{2}\left(T^{b}_{N}(\Phi^{i-1})+T^{b}_{N}(\Phi^{i+1})\right)
\end{equation}
as defined above.

\section{Symmetries for the coupled string dynamics}

In the previous section
we defined $8$ different string dynamics
for each $N$. But are all of these dynamics really different, or do some of them
yield identical or related trajectories $\mathbf{\Phi}_{t}$ in the phase space $S$?\\

Define $\mathbf{P}:S\to S$, $\mathbf{O}:S\to S$ and $\mathbf{E}:S\to S$ by

\begin{equation}
 P^{i}(\mathbf{\Phi})=-\Phi^{i},
\end{equation}
\begin{equation}
 O^{i}(\mathbf{\Phi})=\left\{\begin{array}{ccc}
                              -\Phi^{i}&:&i\mbox{ odd}\\
                               \Phi^{i}&:&i\mbox{ even},
                             \end{array}
                       \right.
\end{equation}
\begin{equation}
 E^{i}(\mathbf{\Phi})=\left\{\begin{array}{ccc}
                              \Phi^{i}&:&i\mbox{ odd}\\
                               -\Phi^{i}&:&i\mbox{ even}.
                             \end{array}
                       \right.
\end{equation}
These operations correspond to discrete $\mathbb{Z}_2$ symmetry
transformations at either all lattice sites, or at odd/even lattice
sites.
 One can easily see that

\begin{eqnarray}
\mathbf{P}\circ\mathbf{P}=\mathbf{E}\circ\mathbf{E}=\mathbf{O}\circ\mathbf{O}
=\mathbf{1},\\
 \mathbf{E}\circ\mathbf{O}=\mathbf{O}\circ\mathbf{E}=\mathbf{P}.
\end{eqnarray}

We are now interested in the state $\mathbf{\Phi}_{t}$ of the chaotic string at a certain time~$t$, which is developed from the initial state $\mathbf{\Phi}_{0}$ by applying $t$ times the function~$\mathbf{f}$:

\begin{equation}
 \mathbf{\Phi}_{t}=\mathbf{f}(\mathbf{\Phi}_{t-1};a)=(\mathbf{f}\circ\mathbf{f}
)(\mathbf{\Phi}_{t-2};a)=\ldots=\mathbf{f}^{t}(\mathbf{\Phi}_{0};a).
\end{equation}

Recall that Tchebyscheff maps of odd order $N_{o}=1,3,...$ are
odd functions of $\Phi$,

\begin{equation}
 T_{N_{o}}(-\Phi)=-T_{N_{o}}(\Phi),
\end{equation}
whereas Tschebyscheff maps of even order $N_{e}=2,4,...$ are even functions of $\Phi$,

\begin{equation}
 T_{N_{e}}(-\Phi)=T_{N_{e}}(\Phi).
\end{equation}
Using the odd/even character of the Tchebyscheff polynomials and the
transformations $\mathbf{P}$, $\mathbf{O}$ and
$\mathbf{E}$, one can derive the following relations (for the
explicit calculations see the Appendix):

\paragraph{Chaotic strings based on odd-$N$ Tchebyscheff
polynomials $(N_{o}=1,3,...)$:}

\begin{eqnarray}
\label{eq:first_relation}
  (\mathbf{f}_{[-N_{o},+,1]})^{t}&=& \left\{\begin{array}{ccc}
                                     (\mathbf{f}_{[N_{o},+,1]})^{t} &:& t\mbox{ even}\\

\mathbf{P}\circ(\mathbf{f}_{[N_{o},+,1]})^{t} &:& t\mbox{ odd},
                                    \end{array}
                        \right.\\
(\mathbf{f}_{[N_{o},-,1]})^{t}&=&
\mathbf{O}\circ(\mathbf{f}_{[N_{o},+,1]})^{t}\circ\mathbf{O},\\
 (\mathbf{f}_{[-N_{o},-,1]})^{t}&=& \left\{\begin{array}{ccc}
                                     \mathbf{O}\circ(\mathbf{f}_{[N_{o},+,1]})^{t}\circ\mathbf{O} &:& t\mbox{ even}\\

\mathbf{E}\circ(\mathbf{f}_{[N_{o},+,1]})^{t}\circ\mathbf{O} &:& t\mbox{ odd},
                                    \end{array}
                        \right.\\
 (\mathbf{f}_{[-N_{o},+,0]})^{t}&=& \left\{\begin{array}{ccc}
                                     \mathbf{O}\circ(\mathbf{f}_{[N_{o},+,0]})^{t}\circ\mathbf{O} &:& t\mbox{ even}\\

\mathbf{E}\circ(\mathbf{f}_{[N_{o},+,0]})^{t}\circ\mathbf{O} &:& t\mbox{ odd},
                                    \end{array}
                 \right.\\
(\mathbf{f}_{[N_{o},-,0]})^{t}&=&
\mathbf{O}\circ(\mathbf{f}_{[N_{o},+,0]})^{t}\circ\mathbf{O},\\
 (\mathbf{f}_{[-N_{o},-,0]})^{t}&=& \left\{\begin{array}{ccc}
                                     (\mathbf{f}_{[N_{o},+,0]})^{t} &:& t\mbox{ even}\\

\mathbf{P}\circ(\mathbf{f}_{[N_{o},+,0]})^{t} &:& t\mbox{ odd}.
                                    \end{array}
                 \right.
\end{eqnarray}

\paragraph{Chaotic strings based on even-$N$ Tchebyscheff
polynomials $(N_{e}=2,4,...)$:}

\begin{eqnarray}
\label{eq:2M_2M} (\mathbf{f}_{[-N_{e},+,1]})^{t}&=&
\mathbf{P}\circ(\mathbf{f}_{[N_{e},+,1]})^{t},\\
 (\mathbf{f}_{[-N_{e},-,1]})^{t}&=&
\mathbf{P}\circ(\mathbf{f}_{[N_{e},-,1]})^{t},\\
 (\mathbf{f}_{[-N_{e},+,0]})^{t}&=&
\mathbf{P}\circ(\mathbf{f}_{[N_{e},+,0]})^{t}\circ\mathbf{P},\\
 (\mathbf{f}_{[-N_{e},-,0]})^{t}&=&
\mathbf{P}\circ(\mathbf{f}_{[N_{e},-,0]})^{t-1}\circ\mathbf{f}_{[N_{e},+,0]}.
\label{eq:last_relation}
\end{eqnarray}
Physically most relevant are the chaotic strings with $N=2$ and
$N=3$. These were used in \cite{Beck:2002, physica-d, dark,
beck-recent} to construct a possible mechanism for fixing and
stabilizing standard model parameters. We come to the conclusion
that the investigation of the former $2\times 8$ different
dynamics can be reduced to the 6~dynamics
$\mathbf{f}_{2A}:=\mathbf{f}_{[2,+,1]}$,
$\mathbf{f}_{2A^{-}}:=\mathbf{f}_{[2,-,1]}$,
$\mathbf{f}_{2B}:=\mathbf{f}_{[2,+,0]}$,
$\mathbf{f}_{2B^{-}}:=\mathbf{f}_{[2,-,0]}$ ,
$\mathbf{f}_{3A}:=\mathbf{f}_{[3,+,1]}$ and
$\mathbf{f}_{3B}:=\mathbf{f}_{[3,+,0]}$, where we have labeled
the dynamics as in \cite{Beck:2002}. All other dynamics yield
trajectories which differ from these six dynamics just by a sign
either at all or at every second lattice site.  But one has to be
careful: Whereas for instance $\mathbf{f}_{[2,+,1]}$ and
$\mathbf{f}_{[-2,+,1]}$ yield the same trajectory up to a sign
for identical initial conditions, the change
$\mathbf{f}_{[2,+,0]}\to \mathbf{f}_{[-2,+,0]}$ is equivalent to a
change of the initial conditions
$\mathbf{\Phi}_{0}\to-\mathbf{\Phi}_{0}$, generally leading to a
different trajectory.

\section{Symmetries of the vacuum energies of chaotic strings}

We now investigate the symmetry behaviour of expectations of
observables associated with the chaotic dynamics.
At this point we consider scalar observables $F:S \to \mathbb{R}$ and
define expectation values as
\begin{eqnarray}
 F_{[N,b,s]}(a)=\left<F(\mathbf{\Phi})\right>&=&\lim_{T\to\infty}\frac{1}{T-T_{0}}\sum_{t=T_{0}}^{T}F(\mathbf{\Phi}_{t}),\nonumber\\
\mbox{with   }\ \ \mathbf{\Phi}_{t}
&=&(\mathbf{f}_{[N,b,s]})^{t}(\mathbf{\Phi}_{0};a).
\end{eqnarray}
The expectation
depends on the underlying dynamics $(N,b,s)$, the coupling~$a$ and the form of
the
observable $F$. Of course, for numerical calculations one will choose a finite
number of iterations $T$ and a finite transient $T_{0}$, getting some small
statistical fluctuations from this. However, note that also for $T\to \infty$
for a given (non-ergodic) dynamics $\mathbf{f}_{[N,s,b]}(\mathbf{\Phi};a)$
the value $F_{[N,b,s]}(a)$ may depend on the initial conditions $\mathbf{\Phi}_{0}$.
In this section we will assume that for the considered dynamics almost all initial conditions
yield the same expectation value -- if we later observe for some $\mathbf{f}_{[N,s,b]}(\mathbf{\Phi};a)$
the contrary, we will state that an expectation for the respective dynamics is
not well-defined. This case corresponds to non-ergodic behaviour
of the CML (see Section~\ref{sec:deviation}).\\

As\ shown in \cite{Beck:2002, physica-d}, there are two physically important
observables for chaotic strings. These are the self energy and
the interaction energy. One defines a formal self-energy potential
$V^{(N)}_{\pm}(\Phi)$ for a chaotic Tchebyscheff map of order $N$ via
\cite{Beck:2002}
\begin{equation}
 \Phi_{t+1}-\Phi_{t}=T_{\pm N}(\Phi_t)=\pm T_{N}(\Phi_t)=-\frac{\partial}{\partial \Phi_t}V^{(N)}_{\pm}(\Phi_t),
\end{equation}
leading to
\begin{eqnarray}
 V_{\pm}^{(2)}(\Phi) &=& \pm \left(-\frac{2}{3}\Phi^{3}+\Phi\right) +\frac{1}{2}\Phi^{2}+C\\
 V_{\pm}^{(3)}(\Phi) &=& \pm \left(-\Phi^{4}+\frac{3}{2}\Phi^{2}\right) +\frac{1}{2}\Phi^{2}+C.
\end{eqnarray}
Here $C$ is an arbitrary additive constant. Similarly one
can define a formal interaction potential $aW_{\pm}(\Phi^{i},\Phi^{j})$
between neighbouring sites $i$ and $j$ by
\begin{equation}
W_{s}(\Phi^{i},\Phi^{j})=\frac{1}{4}(\Phi^{i}-s\Phi^{j})^{2}+C,
\end{equation}
which delivers the building block of the diffusive (s=``$+$'') or
anti-diffusive (s=``$-$'') interaction via
\begin{equation}
 -\frac{\partial}{\partial
\Phi^{i}}W_{s}(\Phi^{i},\Phi^{j})=-\frac{1}{2}(\Phi^{i}-s\Phi^{j}).
\end{equation}
In \cite{Beck:2002} the additive constant~$C$ was chosen as
\begin{equation}
 C=-\frac{1}{2}\left<\Phi^{2}\right> ,\label{34}
\end{equation}
by which we obtain the expectation value of $V_{\pm}$ for the dynamics
$\mathbf{f}_{[N,b,s]}$ at given lattice site $i$ as
\begin{eqnarray}
V^{i}_{[\pm 2,s,b]}(a) &=&
\pm\left<-\frac{2}{3}(\Phi^{i})^{3}+(\Phi^{i})\right>,\\
V^{i}_{[\pm 3,s,b]}(a) &=&
\pm\left<-(\Phi^{i})^{4}+\frac{3}{2}(\Phi^{i})^{2}\right>.
\end{eqnarray}
This expectation is a function of the coupling $a$. Similary, for
the expectation value of the interaction energy between a
neighbouring pair of lattice sites~$i,j$ we obtain
\begin{equation}
 W^{i,j}_{[N,s,b]}(a) = -s\frac{1}{2}\left<\Phi^{i}\Phi^{j}\right>
\end{equation}
where the variable $s$ accounts for the definition that in case
of (anti-) diffusive coupling the interaction energy gets a
(positive) negative sign. The expectations of self-energy and
interaction energy are physically interpreted as two different
types of vacuum energies (see \cite{Beck:2002} for details).

In the following we assume that both types of vacuum energies do
not depend on the lattice indices where they are evaluated and
omit the indices $i,j$. This requested independence is closely
related to the independence on the initial conditions and the
ergodicity of the system. Whether the assumption is true or not in
the various coupling parameter regions will be investigated in
more detail in section 5.

As shown in section 3, all 16 possible combinations $[\pm N,b,s]$
for $N=2,3$ basically lead to six different dynamics. We will now
examine the effect of switching between positive and negative
Tchebyscheff polynomials ($N\to -N$) and between diffusive and
anti-diffusive coupling $(s=1)\to (s=-1)$ for both types of vacuum
energies.\footnote{Note that analogous relations hold for self-
and interaction energies of chaotic strings defined for arbitrary
$N$. Of relevance is only the fact whether $N$ is even or odd.}

\paragraph{Transformation $N\to-N$:}
For all chaotic string dynamics with $N=2$ the vacuum energies
transform as

\begin{eqnarray}
V_{[+2,s,b]}(a) &\to& V_{[-2,s,b]}(a)=V_{[2,s,b]}(a),\\
W_{[+2,s,b]}(a) &\to& W_{[-2,s,b]}(a)=W_{[2,s,b]}(a).
\end{eqnarray}

In contrast, the $N=3$ dynamics show a different behaviour:

\begin{eqnarray}
 V_{[+3,s,b]}(a) &\to& V_{[-3,s,b]}(a)=-V_{[3,s,b]}(a),\\
 W_{[+3,s,1]}(a) &\to& W_{[-3,s,1]}(a)=W_{[3,s,1]}(a),\\
 W_{[+3,s,0]}(a) &\to& W_{[-3,s,0]}(a)=-W_{[3,s,0](a)}.
\end{eqnarray}

Note that the above well-defined symmetry behaviour was only
achieved due to our special choice of the additive constant $C$
in Eq.~(\ref{34}). Other choices would not make the problem
symmetric under the transformation $N \to -N$.
We see that the theory of chaotic strings very much depends on invariance
under suitable discrete $\mathbb{Z}_2$ group
transformations. As emphasized in \cite{Beck:2002},
physically relevant observables should have
a well-defined transformation behaviour under this operation.
This can be seeen in analogy to the fact that in gauge theories
physically relevant observables should be gauge invariant.
For chaotic strings one has the
unique possibility to fix additive constants of vacuum energies
(relevant in gravitational theories) by discrete symmetry
considerations \cite{Beck:2002}.

\paragraph{Transformation $(s=1)\to (s=-1)$:}
Whereas for chaotic string dynamics with $N=2$ this transformation
yields a completely different behaviour, for $N=3$ one obtains
\begin{eqnarray}
 V_{[\pm3,+,b]}(a) &\to& V_{[\pm 3,-,b]}(a)=V_{[\pm 3,+,b]}(a),\\
 W_{[\pm 3,+,b]}(a) &\to& W_{[\pm 3,-,b]}(a)=W_{[\pm 3,+,b]}(a).
\end{eqnarray}

\section{Dependence of expectations on initial values and lattice position}
\label{sec:deviation}

If the CML exhibits ergodic behaviour, then expectations of
observables will neither depend on lattice position nor on the
initial values (up to a set of measure 0). To investigate this,
we define for the six interesting dynamics
$3A,3B,2A,2B,2A^{-},2B^{-}$ the following measures of
inhomogeneity:
\begin{eqnarray}
 \sigma_{\mathrm{init}}(W;a)&=& \left<\sqrt{
\left<(W^{i,j}(a))^{2}\right>_{\mathbf{\Phi}_{0}}-\left<W^{i,j}(a)\right>^{2}_{
\mathbf{\Phi}_{0}}}\right>_{\mathrm{lattice}},\\
 \sigma_{\mathrm{lattice}}(W;a)&=& \left<\sqrt{
\left<(W^{i,j}(a))^{2}\right>_{\mathrm{lattice}}-\left<W^{i,j}(a)\right>^{2}_{
\mathrm {lattice}}}\right>_{\mathbf{\Phi}_{0}},
\end{eqnarray}
where $\left<W\right>_{\mathrm{lattice}}$ refers to an average
of an observable $W$ taken over the lattice and $\left<W\right>_{\mathbf{\Phi}_{0}}$
to an average taken over an ensemble of initial values. Let us explain
these definitions: To calculate $\sigma_{\mathrm{init}}$,
we pick a dynamics, for instance $3B$, and
some coupling $a$ and calculate for an ensemble of initial values the
local observable, in our case here chosen to be
the interaction energy $W^{i,j}(a)$ for neighbouring
lattice points (i.e. $j=i+1$). Then we determine the
standard deviation of this quantity with respect to the
ensemble of initial values.
Finally, we take the average over the lattice, getting a measure
of the inhomogeneity of the interaction energy with respect to
different initial values. To calculate the quantity $\sigma_{\mathrm{lattice}}$
we perform the above averaging procedures in reverse order.

Fig.~1 shows the result of our numerical calculation
for the six interesting string dynamics
as a function of the coupling $a$.
Apparently both $\sigma_{\mathrm{init}}(W;a)$ and
$\sigma_{\mathrm{lattice}}(W;a)$ show nearly identical behaviour
and are non-vanishing for large ranges of the coupling parameter
$a$, thus indicating non-ergodic behaviour.

\begin{figure}
\begin{center}
\subfigure[$2A$]{\includegraphics[scale=0.4]{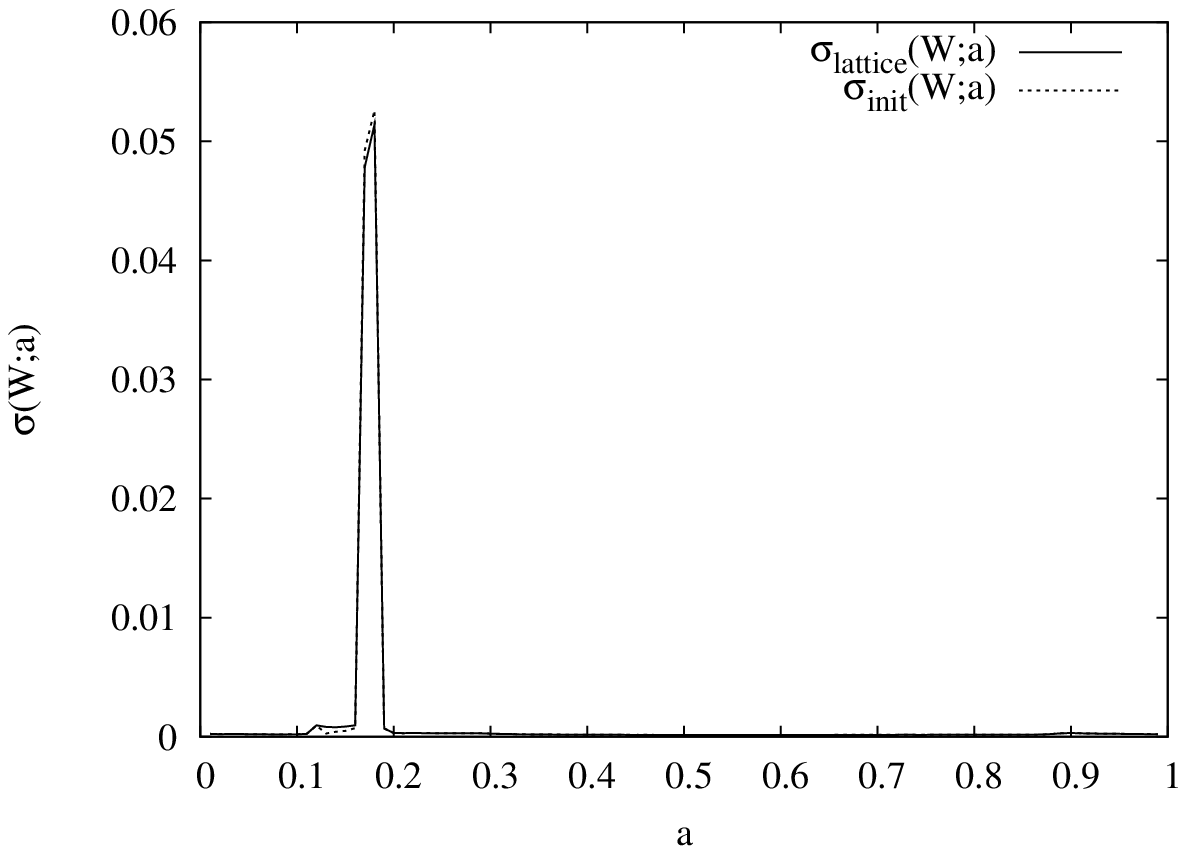}}
\subfigure[$2B$]{\includegraphics[scale=0.4]{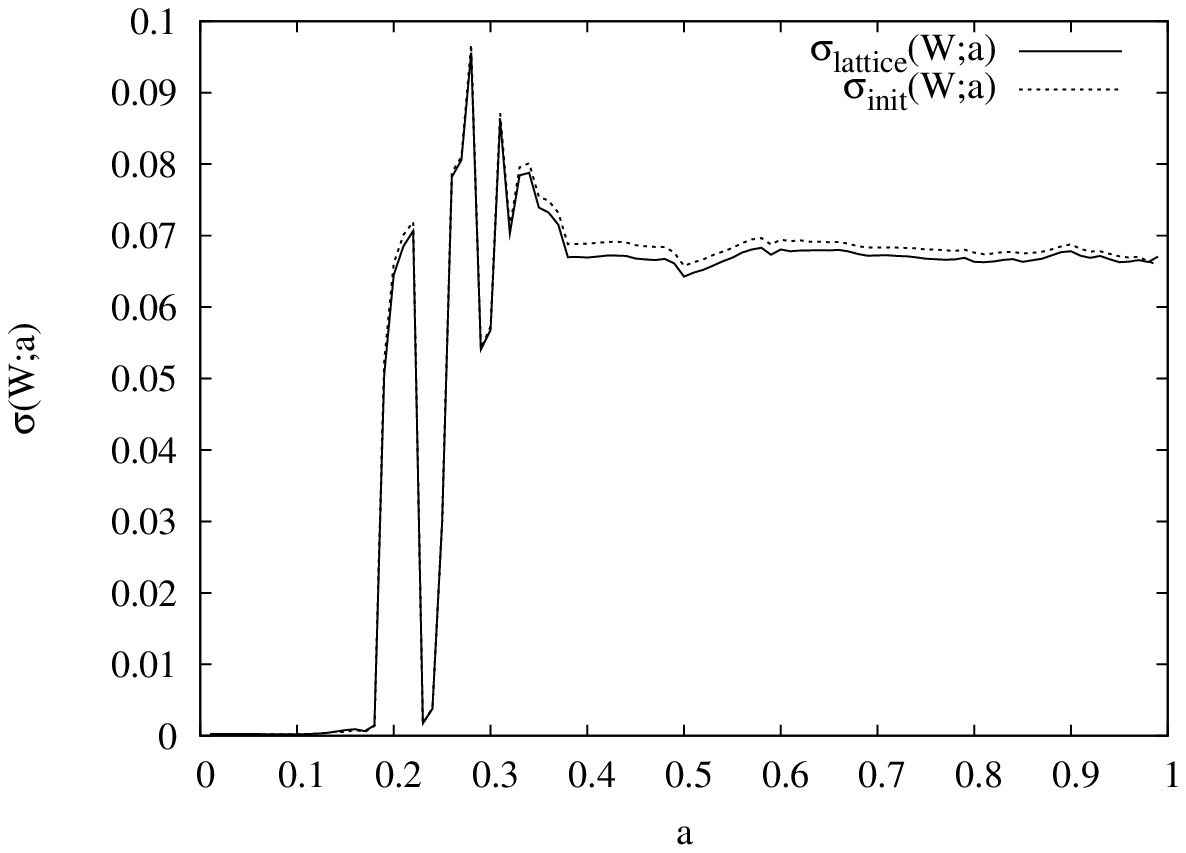}}
\subfigure[$2A^{-}$]{\includegraphics[scale=0.4]{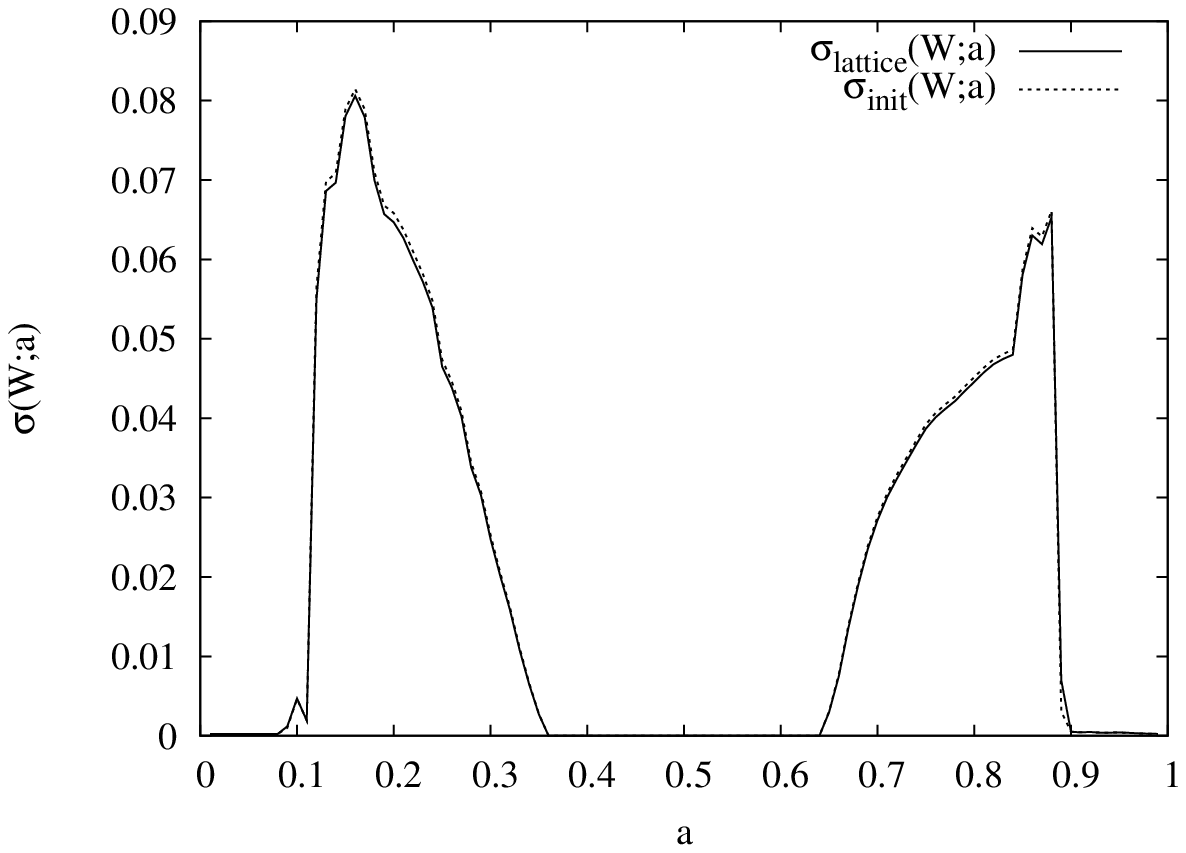}}
\subfigure[$2B^{-}$]{\includegraphics[scale=0.4]{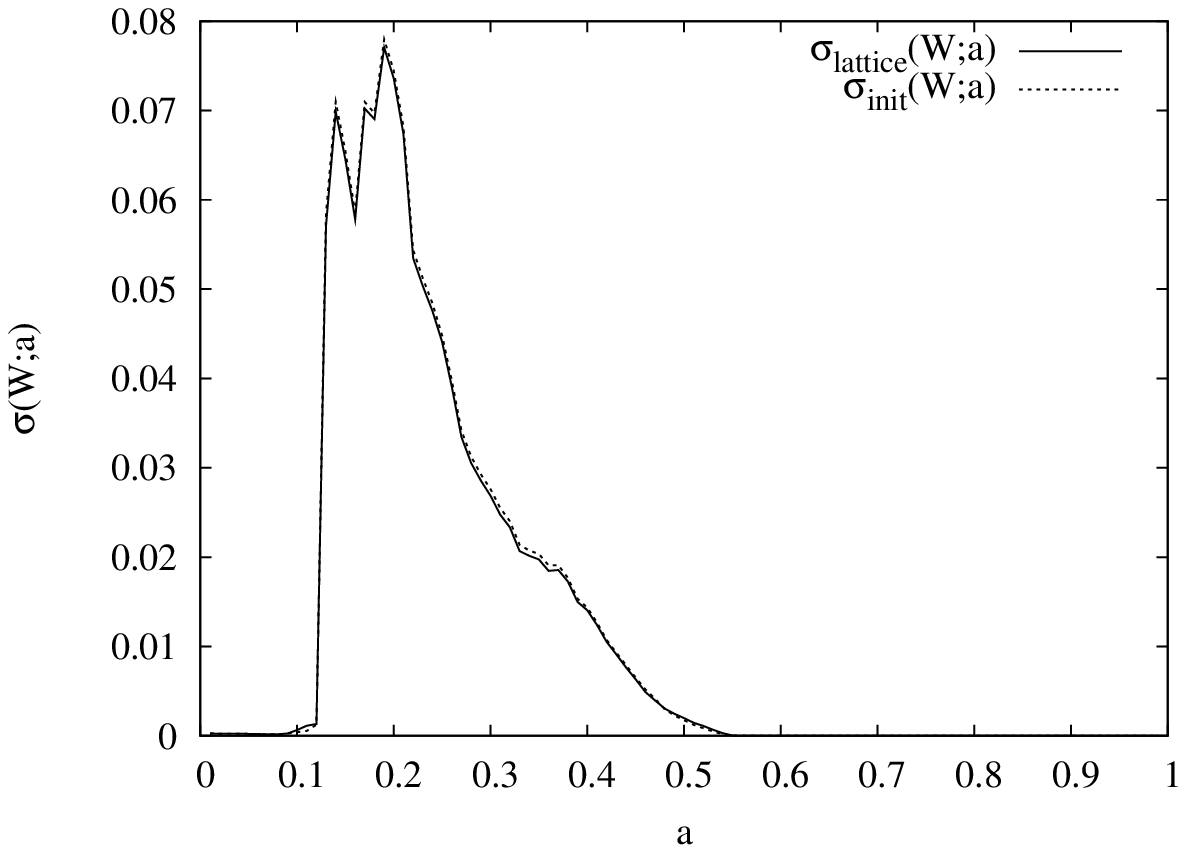}}
\subfigure[$3A$]{\includegraphics[scale=0.4]{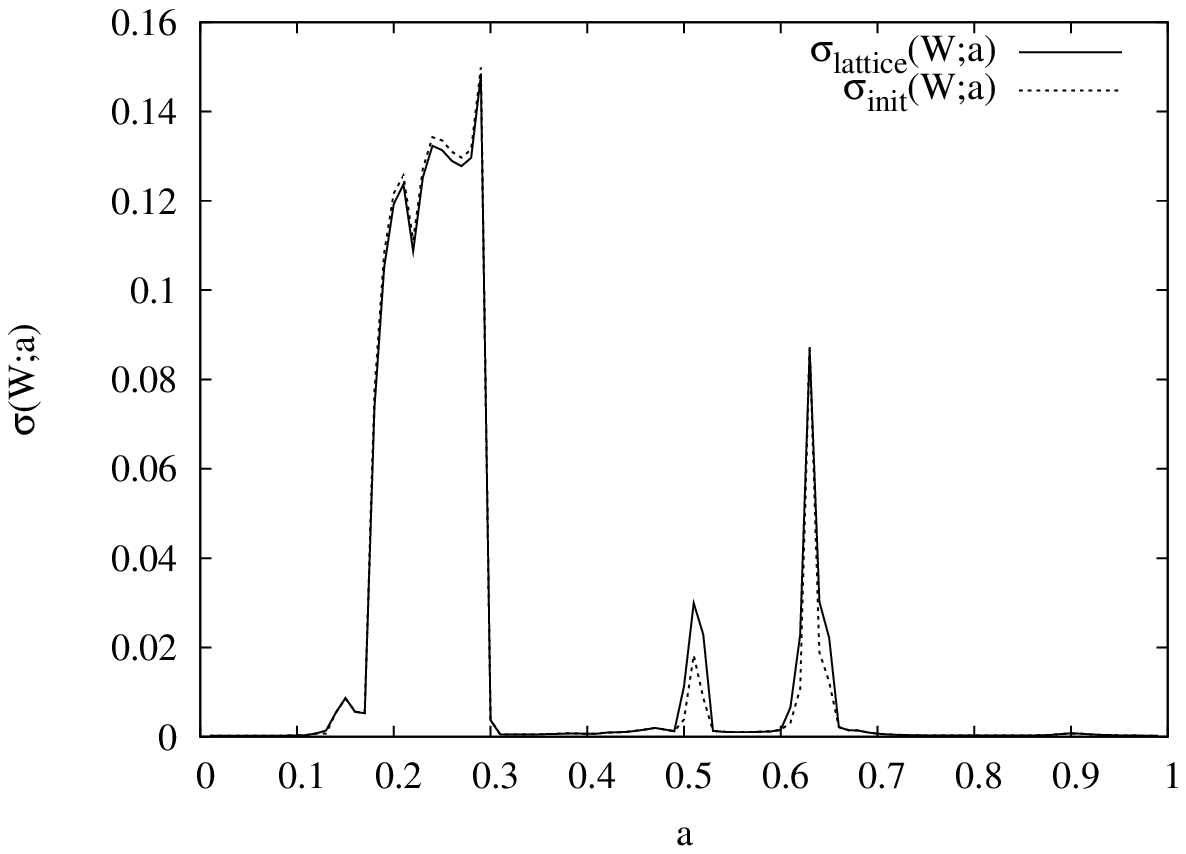}}
\subfigure[$3B$]{\includegraphics[scale=0.4]{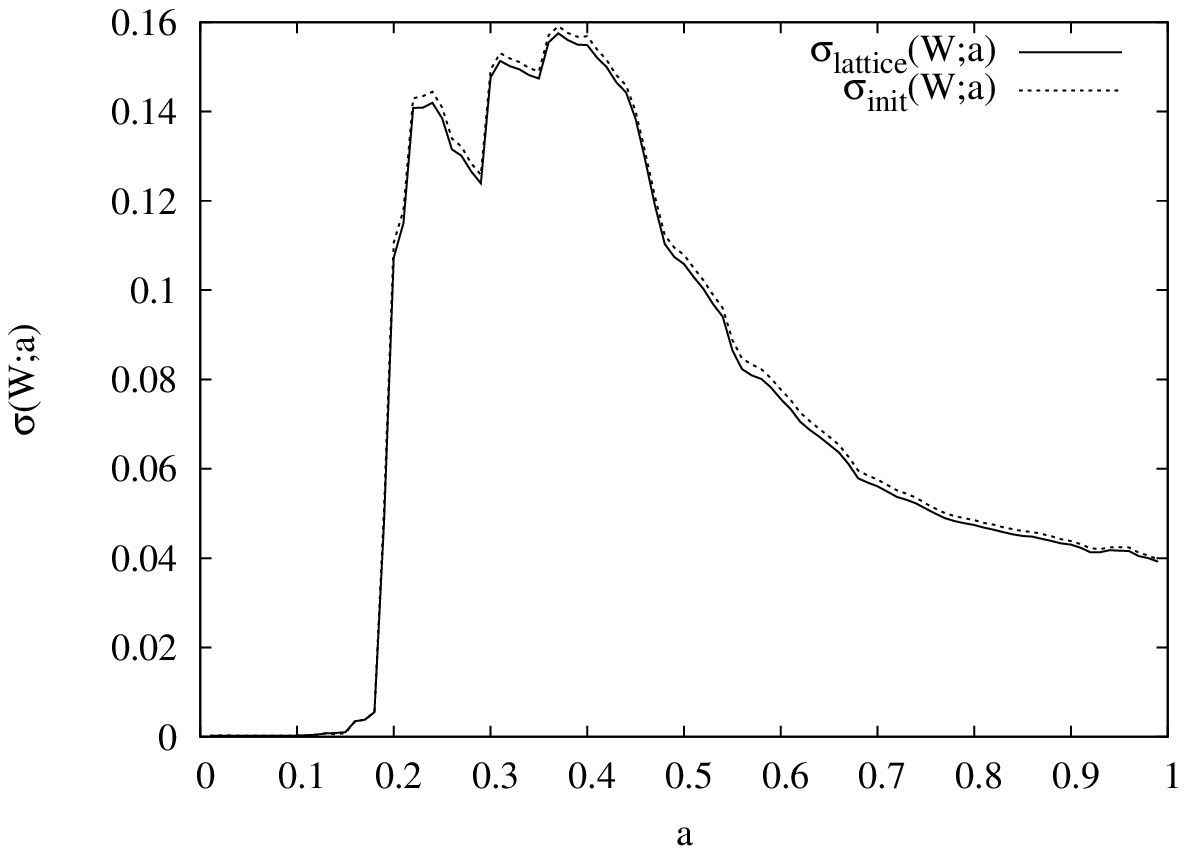}}
\end{center}
\caption{Inhomogeneity measures $\sigma_{\mathrm{init}}(W;a)$ and
$\sigma_{\mathrm{lattice}}(W;a)$ for the six different chaotic
strings. For the numerical calculation we used  50  different
initial conditions, a lattice size $J=500$, $T=10^7$ iterations
and a transient of $5\times 10^4$.} \label{fig:deviation}
\end{figure}

We may also look at the joined inhomogeneity measure
$\sigma(W;a)$ given by
\begin{equation}
 \sigma(W;a)= \sqrt{
\left<(W^{i,j}(a))^{2}\right>_{\mathrm{lattice},\Phi_{0}}-\left<W^{i,j}
(a)\right>^
{2}_{\mathrm{lattice},\Phi_{0}}},
\end{equation}
where the standard deviation is now simulataneously determined from an average
over all initial values and all lattice sites.
For a finite number of iterations $T$ and for finite lattice
sizes statistical disorder will
prevent a completely vanishing standard deviation. As a criterium
to distinguish between an inhomogeneous non-ergodic dynamics and a
homogeneous ergodic dynamics with a nonzero $\sigma(W;a)$ due to
finite-iteration time statistical fluctuations, we determined
$\sigma(W;a)$ as a function of the number of iterations. For
statistical fluctuations we expect
\begin{equation}
 \sigma(W;a|T)\propto T^{-\frac{1}{2}}
\end{equation}
(due to the Central Limit Theorem)
whereas for an inhomogeneous (non-ergodic) dynamics
we expect
\begin{equation}
 \sigma(W;a|T)\approx \mbox{constant} \label{49}
\end{equation}
due to
very slow decay of correlations and different attractors
in the system.
This behaviour is verified in Fig.~2 for some important examples of string dynamics
and coupling constants.

In \cite{Beck:2002} the smallest stable zeros of the interaction
energy, i.e. values $a^*$ with $W_{[N,s,b]}(a^*)=0$ and
$W'_{[N,s,b]}(a^{*})<0$, were used to fix standard model coupling
constants from first principles. The idea advocated in
\cite{Beck:2002} is that standard model coupling constants are
fixed as states of strongest random properties of chaotic
strings, distinguished by a vanishing nearest-neighbour
correlation.
 Interesting enough, in \cite{Beck:2002} six stable zeros were found for which a
direct standard model interpretation was possible. These were the
smallest stable zeros 0.0008164, 0.007304, 0.001801, 0.01755 of
the $3A$ and $3B$ dynamics, which could be associated with
electroweak interactions strengths at the lightest fermionic mass
scales, and the smallest stable zeros 0.1201 and 0.09537 of the
$2A$ and $2B^-$ dynamics, which could be associated with strong
interaction strengths at the $W$ and Higgs mass scale,
respectively. Two additional stable zeros were found, namely the
smallest stable zero $0.1758$ of the $2A^-$ string and the
smallest stable zero $0.3145$ of the $2B$ string, for which the
physical interpretation was by far less obvious. In Fig.~2 the
quantity  $\sigma(W;a|T)$ is displayed for all these zeros.
Interesting enough, we see from this figure that the two zeros
that could not be interpreted in a straightforward way in terms
of standard model interaction strengths apparently correspond to
nonergodic states with non-converging behaviour of type
(\ref{49}). These zeros should thus be discarded from the
analysis. On the other hand, the other zeros which do have
physical meaning in \cite{Beck:2002} are observed to correspond
to ergodic states with well-defined convergence behaviour.
Apparently, physically relevant states of the vacuum should be
associated with ergodic behaviour.


\begin{figure}
\begin{center}
\includegraphics{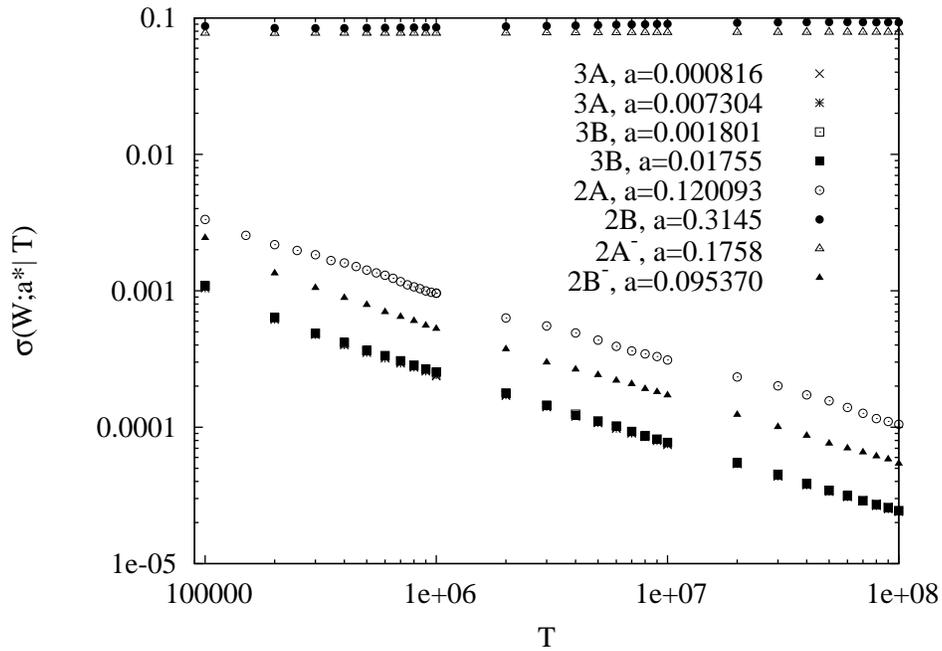}
\end{center}
\label{fig:zeros} \caption{The time-dependent standard deviation
$\sigma(W;a|T)$ for eight distinguished values $a^{*}$ with
$W(a^{*})=0$. The averages are taken over lattices with $J=500$
sites and $10$ different initial conditions. A transient
$T_{0}=5\times10^{4}$ has been discarded from the calculation.}
\end{figure}


\newpage

\begin{appendix}
\section{Appendix: Derivation of Eq.~(\ref{eq:first_relation}) -- Eq.~(\ref{eq:last_relation})}

\paragraph{Chaotic strings derived from Tchebyscheff polynoms with odd order~$N_{o}=1,3,...$:}
Remember that $T_{N_{o}}(-\Phi)=-T_{N_{o}}(\Phi)$.

\begin{eqnarray}
  f_{[-N_{o},+,1]}^{i}(\mathbf{\Phi};a) &=& (1-a)(-T_{N_{o}}(\Phi^{i}))+\frac{a}{2}\left\{(-T_{N_{o}}(\Phi^{i-1}))+(-T_{N_{o}}(\Phi^{i+1}))\right\}\nonumber\\
&=& -\left\{(1-a)T_{N_{o}}(\Phi^{i})+\frac{a}{2}\left(T_{N_{o}}(\Phi^{i-1})+T_{N_{o}}(\Phi^{i+1}\right)\right\}\nonumber\\
&=& (1-a)T_{N_{o}}(-\Phi^{i})+\frac{a}{2}\left(T_{N_{o}}(-\Phi^{i-1})+T_{N_{o}}(-\Phi^{i+1}\right)\nonumber\\
&=& -f_{[N_{o},+,1]}^{i}(\mathbf{\Phi};a) = f_{[N_{o},+,1]}^{i}(-\mathbf{\Phi};a)\nonumber\\
\rightarrow (\mathbf{f}_{[-N_{o},+,1]})^{t}&=&
\left\{\begin{array}{ccc}
                                    (\mathbf{f}_{[N_{o},+,1]}\circ\mathbf{P}\circ\mathbf{P}\circ\mathbf{f}_{[N_{o},+,1]})^{t/2} &:& t\mbox{ even}\\
                                    \mathbf{P}\circ\mathbf{f}_{[N_{o},+,1]}\circ(\mathbf{f}_{[N_{o},+,1]}\circ\mathbf{P}\circ\mathbf{P}\circ\mathbf{f}_{[N_{o},+,1]})^{(t-1)/2} &:& t\mbox{ odd}
                                    \end{array}
                        \right.\nonumber\\
&=&
\left\{\begin{array}{ccc}
                                    (\mathbf{f}_{[N_{o},+,1]}^{t} &:& t\mbox{ even}\\
                                    \mathbf{P}\circ(\mathbf{f}_{[N_{o},+,1]})^{t} &:& t\mbox{ odd}
                                    \end{array}
                        \right.
\end{eqnarray}

\begin{equation}
  f_{[N_{o},-,1]}^{i}(\mathbf{\Phi};a) = (1-a)(T_{N_{o}}(\Phi^{i}))-\frac{a}{2}\left\{(T_{N_{o}}(\Phi^{i-1}))+(T_{N_{o}}(\Phi^{i+1}))\right\}\nonumber
\end{equation}

For odd lattice sites $i$ one has

\begin{eqnarray}
 f_{[N_{o},-,1]}^{i}(\mathbf{O}\circ\mathbf{\Phi};a) &=&
(1-a)T_{N_{o}}(-\Phi^{i})-\frac{a}{2}\left\{T_{N_{o}}(\Phi^{i-1})+T_{N_{o}}(\Phi^{i+1})\right\}\nonumber\\
O^{i}((\mathbf{f}_{[N_{o},-,1]}\circ\mathbf{O})(\mathbf{\Phi};a)) &=&
-\left\{(1-a)T_{N_{o}}(-\Phi^{i})-\frac{a}{2}\left\{T_{N_{o}}(\Phi^{i-1})+T_{N_{o}}(\Phi^{i+1})\right\}\right\}\nonumber\\
&=&(1-a)T_{N_{o}}(\Phi^{i})+\frac{a}{2}\left\{T_{N_{o}}(\Phi^{i-1})+T_{N_{o}}(\Phi^{i+1})\right\}\nonumber
\end{eqnarray}

For even lattice sites $i$ one has

\begin{eqnarray}
 f_{[N_{o},-,1]}^{i}(\mathbf{O}\circ\mathbf{\Phi};a) &=&
(1-a)T_{N_{o}}(\Phi^{i})-\frac{a}{2}\left\{T_{N_{o}}(-\Phi^{i-1})+T_{N_{o}}(-\Phi^{i+1})\right\}\nonumber\\
O^{i}((\mathbf{f}_{[N_{o},-,1]}\circ\mathbf{O})(\mathbf{\Phi};a)) &=&
(1-a)T_{N_{o}}(\Phi^{i})-\frac{a}{2}\left\{T_{N_{o}}(-\Phi^{i-1})+T_{N_{o}}(-\Phi^{i+1})\right\}\nonumber\\
&=&(1-a)T_{N_{o}}(\Phi^{i})+\frac{a}{2}\left\{T_{N_{o}}(\Phi^{i-1})+T_{N_{o}}(\Phi^{i+1})\right\}\nonumber
\end{eqnarray}

This leads to

\begin{eqnarray}
\mathbf{O}\circ\mathbf{f}_{[N_{o},-,1]}\circ\mathbf{O}&=& \mathbf{f}_{[N_{o},+,1]}\nonumber\\
\label{eq:f_3minus1}
\rightarrow (\mathbf{f}_{[N_{o},-,1]})^{t}&=& \mathbf{O}\circ(\mathbf{f}_{[N_{o},+,1]})^{t}\circ\mathbf{O}
\end{eqnarray}

\begin{eqnarray}
  f_{[-N_{o},-,1]}^{i}(\mathbf{\Phi};a) &=& (1-a)(-T_{N_{o}}(\Phi^{i}))-\frac{a}{2}\left\{-T_{N_{o}}(\Phi^{i-1})-T_{N_{o}}(\Phi^{i+1})\right\}\nonumber\\
&=&
-\left\{(1-a)(T_{N_{o}}(\Phi^{i}))-\frac{a}{2}\left\{T_{N_{o}}(\Phi^{i-1})+T_{N_{o}}(\Phi^{i+1}))\right\}\right\}\nonumber\\
&=&
(1-a)(T_{N_{o}}(-\Phi^{i}))-\frac{a}{2}\left\{T_{N_{o}}(-\Phi^{i-1})+T_{N_{o}}(-\Phi^{i+1})\right\}\nonumber\\
\rightarrow \mathbf{f}_{[-N_{o},-,1]}&=& \mathbf{P}\circ\mathbf{f}_{[N_{o},-,1]}=\mathbf{f}_{[N_{o},-,1]}\circ\mathbf{P}\nonumber\\
\label{eq:f_m3minusm1}
\rightarrow (\mathbf{f}_{[-N_{o},-,1]})^{t}&=&
\left\{\begin{array}{ccc}
                                    \mathbf{O}\circ(\mathbf{f}_{[N_{o},+,1]})^{t}\circ\mathbf{O} &:& t\mbox{ even}\\
                                    \mathbf{E}\circ(\mathbf{f}_{[N_{o},+,1]})^{t}\circ\mathbf{O} &:& t\mbox{ odd}
                                    \end{array}
                        \right.
\end{eqnarray}

In the  last step $\mathbf{P}\circ\mathbf{O}=\mathbf{E}$ was used.

\begin{equation}
  f_{[N_{o},-,0]}^{i}(\mathbf{\Phi};a) = (1-a)(T_{N_{o}}(\Phi^{i}))-\frac{a}{2}\left\{\Phi^{i}+\Phi^{i}\right\}\nonumber
\end{equation}

Repeating the steps which took us to Eq.~(\ref{eq:f_3minus1}), we
arrive at

\begin{equation}
(\mathbf{f}_{[N_{o},-,0]})^{t}= \mathbf{O}\circ(\mathbf{f}_{[N_{o},+,0]})^{t}\circ\mathbf{O}
\end{equation}

\begin{eqnarray}
  f_{[-N_{o},+,0]}^{i}(\mathbf{\Phi};a) &=&
(1-a)(-T_{N_{o}}(\Phi^{i}))+\frac{a}{2}\left\{\Phi^{i}+\Phi^{i}\right\}\nonumber\\
&=& -\left\{(1-a)T_{N_{o}}(\Phi^{i})-\frac{a}{2}\left(\Phi^{i}+\Phi^{i}\right)\right\}\nonumber\\
&=& (1-a)T_{N_{o}}(-\Phi^{i})-\frac{a}{2}\left(-\Phi^{i}-\Phi^{i}\right)\nonumber\\
\rightarrow \mathbf{f}_{[-N_{o},+,0]}&=& \mathbf{P}\circ\mathbf{f}_{[N_{o},-,0]}=\mathbf{f}_{[N_{o},-,0]}\circ\mathbf{P}\nonumber\\
\label{eq:f_m3minusm1}
\rightarrow (\mathbf{f}_{[-N_{o},+,0]})^{t}&=&
\left\{\begin{array}{ccc}
                                    \mathbf{O}\circ(\mathbf{f}_{[N_{o},+,0]})^{t}\circ\mathbf{O} &:& t\mbox{ even}\\
                                    \mathbf{E}\circ(\mathbf{f}_{[N_{o},+,0]})^{t}\circ\mathbf{O} &:& t\mbox{ odd}
                                    \end{array}
                        \right.
\end{eqnarray}

\begin{eqnarray}
  f_{[-N_{o},-,0]}^{i}(\mathbf{\Phi};a) &=&
(1-a)(-T_{N_{o}}(\Phi^{i}))-\frac{a}{2}\left\{\Phi^{i}+\Phi^{i}\right\}\nonumber\\
&=& -\left\{(1-a)T_{N_{o}}(\Phi^{i})+\frac{a}{2}\left(\Phi^{i}+\Phi^{i}\right)\right\}\nonumber\\
&=& (1-a)T_{N_{o}}(-\Phi^{i})+\frac{a}{2}\left(-\Phi^{i}-\Phi^{i}\right)\nonumber\\
\rightarrow \mathbf{f}_{[-N_{o},-,0]}&=& \mathbf{P}\circ\mathbf{f}_{[N_{o},+,0]}=\mathbf{f}_{[N_{o},+,0]}\circ\mathbf{P}\nonumber\\
\rightarrow (\mathbf{f}_{[-N_{o},-,0]})^{t}&=&
\left\{\begin{array}{ccc}
                                    (\mathbf{f}_{[N_{o},+,0]})^{t} &:& t\mbox{ even}\\
                                    \mathbf{P}\circ(\mathbf{f}_{[N_{o},+,0]})^{t} &:& t\mbox{ odd}
                                    \end{array}
                        \right.
\end{eqnarray}

 \paragraph{Chaotic strings derived from Tchebyscheff polynoms with even order~$N_{e}$:}

Remember that $T_{N_{e}}(-\Phi)=T_{N_{e}}(\Phi)$.

\begin{eqnarray}
  f_{[-N_{e},+,1]}^{i}(\mathbf{\Phi};a) &=&
(1-a)(-T_{N_{e}}(\Phi^{i}))+\frac{a}{2}\left\{(-T_{N_{e}}(\Phi^{i-1}))+(-T_{
N_{e}}(\Phi^{i+1}))\right\}\nonumber\\
&=& -\left\{(1-a)T_{N_{e}}(\Phi^{i})+\frac{a}{2}\left(T_{N_{e}}(\Phi^{i-1})+T_{N_{e}}(\Phi^{i+1}\right)\right\}\nonumber\\
&=& -f_{[N_{e},+,1]}^{i}(\mathbf{\Phi};a) = -f_{[N_{e},+,1]}^{i}(-\mathbf{\Phi};a)\nonumber\\
\rightarrow (\mathbf{f}_{[-N_{e},+,1]})^{t}&=& (\mathbf{P}\circ\mathbf{f}_{[N_{e},+,1]}\circ\mathbf{P})^{t-1}\circ\mathbf{P}\circ\mathbf{f}_{[N_{e},+,1]}\nonumber\\
&=& \mathbf{P}\circ(\mathbf{f}_{[N_{e},+,1]}\circ\mathbf{P}\circ\mathbf{P})^{t-1}\circ\mathbf{f}_{[N_{e},+,1]}\nonumber\\
&=& \mathbf{P}\circ(\mathbf{f}_{[N_{e},+,1]})^{t}
\end{eqnarray}

\begin{eqnarray}
  f_{[-N_{e},-,1]}^{i}(\mathbf{\Phi};a) &=& (1-a)(-T_{N_{e}}(\Phi^{i}))-\frac{a}{2}\left\{(-T_{N_{e}}(\Phi^{i-1}))+(-T_{N_{e}}(\Phi^{i+1}))\right\}\nonumber\\
&=&
-\left\{(1-a)T_{N_{e}}(\Phi^{i})-\frac{a}{2}\left(T_{N_{e}}(\Phi^{i-1})+T_{N_
{e}}(\Phi^{i+1}\right)\right\}\nonumber\\
&=& -f_{[N_{e},-,1]}^{i}(\mathbf{\Phi};a) = -f_{[N_{e},-,1]}^{i}(-\mathbf{\Phi};a)\nonumber\\
\rightarrow (\mathbf{f}_{[-N_{e},-,1]})^{t}&=& \mathbf{P}\circ(\mathbf{f}_{[N_{e},-,1]})^{t}
\end{eqnarray}

\begin{eqnarray}
  f_{[-N_{e},+,0]}^{i}(\mathbf{\Phi};a) &=& (1-a)(-T_{N_{e}}(\Phi^{i}))+\frac{a}{2}\left\{\Phi^{i-1}+\Phi^{i+1}\right\}\nonumber\\
&=& -\left\{(1-a)T_{N_{e}}(-\Phi^{i})+\frac{a}{2}\left(-\Phi^{i-1}-\Phi^{i+1}\right)\right\}\nonumber\\
&=& -f_{[N_{e},+,0]}^{i}(\mathbf{-\Phi};a)\nonumber\\
\rightarrow (\mathbf{f}_{[-N_{e},+,0]})^{t}&=& (\mathbf{P}\circ\mathbf{f}_{[N_{e},+,0]}\circ\mathbf{P})^{t}\nonumber\\
&=&
 \mathbf{P}\circ(\mathbf{f}_{[N_{e},+,0]}\circ\mathbf{P}\circ\mathbf{P})^{t-1}\circ\mathbf{f}_{[N_{e},+,0]}\circ\mathbf{P}\nonumber\\
&=& \mathbf{P}\circ(\mathbf{f}_{[N_{e},+,0]})^{t}\circ\mathbf{P}
\end{eqnarray}

\begin{eqnarray}
  f_{[-N_{e},-,0]}^{i}(\mathbf{\Phi};a) &=& (1-a)(-T_{N_{e}}(\Phi^{i}))-\frac{a}{2}\left\{\Phi^{i-1}+\Phi^{i+1}\right\}\nonumber\\
&=& -\left\{(1-a)T_{N_{e}}(\Phi^{i})+\frac{a}{2}\left(\Phi^{i-1}+\Phi^{i+1}\right)\right\}\nonumber\\
&=& -f_{[N_{e},+,0]}^{i}(\mathbf{\Phi};a)\nonumber\\
f_{[N_{e},+,0]}^{i}(\mathbf{-\Phi};a) &=& (1-a)(T_{N_{e}}(-\Phi^{i}))+\frac{a}{2}\left\{-\Phi^{i-1}-\Phi^{i+1}\right\}\nonumber\\
&=& (1-a)(T_{N_{e}}(\Phi^{i}))-\frac{a}{2}\left\{\Phi^{i-1}+\Phi^{i+1}\right\}\nonumber\\
&=& f_{[N_{e},-,0]}^{i}(\mathbf{\Phi};a)\nonumber\\
\rightarrow (\mathbf{f}_{[-N_{e},-,0]})^{t}&=& (\mathbf{P}\circ\mathbf{f}_{[N_{e},+,0]})^{t}\nonumber\\
&=& \mathbf{P}\circ(\mathbf{f}_{[N_{e},+,0]}\circ\mathbf{P})^{t-1}\circ\mathbf{f}_{[N_{e},+,0]}\nonumber\\
&=& \mathbf{P}\circ(\mathbf{f}_{[N_{e},-,0]})^{t-1}\circ\mathbf{f}_{[N_{e},+,0]}
\end{eqnarray}

\end{appendix}

\end{document}